\documentclass[a4paper,twocolumn,11pt]{quantumarticle}
\pdfoutput=1

\usepackage[utf8]{inputenc}
\usepackage[english]{babel}
\usepackage[T1]{fontenc}

\usepackage{amsmath,amssymb,amsfonts,mathrsfs}

\usepackage{hyperref}
\hypersetup{breaklinks=true} 

\usepackage{tikz}
\usetikzlibrary{decorations.pathmorphing}

\begin{document}
	
\title{The lifetime of white hole remnants is $M^5$}

\author{Pierre Martin-Dussaud}
\email{martindussaud[at]gmail[dot]com}
\affiliation{Basic Research Community for Physics (BRCP) \\ Quantum Information Structure of Spacetime (QISS)}

\date{ \small April 2025}

\begin{abstract}
	\noindent The \textit{black-to-white hole} scenario is a well-motivated proposal from non-perturbative quantum gravity concerning the final stage of black hole evaporation. In this framework, a black hole of initial mass \( M \) undergoes Hawking evaporation over a timescale of order \( M^3 \) before tunneling into a long-lived white hole. Previous estimates suggest that the lifetime of the resulting white hole scales as \( M^4 \)~\cite{bianchi2018}. In this short note, I argue that such estimates neglect key aspects of the white hole's internal dynamics, and I demonstrate that a more consistent timescale for the white hole lifetime is of order \( M^5 \).
\end{abstract}

\maketitle

\section{Black holes could tunnel into white hole remnants at the end of Hawking evaporation}

In classical general relativity, a collapsing star that crosses its Schwarzschild radius is compelled to undergo complete gravitational collapse, as no known form of matter can resist compression within a black hole. However, quantum gravity suggests that this picture is incomplete, potentially resolving the singularity. Several models propose that a quantum black hole may undergo a tunneling process into a white hole.

The earliest seeds of the black-to-white hole scenario can be found in a 1979 paper by Frolov and Vilkovisky~\cite{frolov1979}. Using an effective theory incorporating first-order quantum corrections to the Einstein-Hilbert action, they studied the dynamics of gravitational collapse. In the case of a spherically symmetric collapse of a null shell of matter, they found that no singularity forms; instead, when the shell reaches \( r = 0 \), it reverses direction and expands outward.

In 2001, Hájíček and Kiefer reached similar conclusions using a reduced quantization method~\cite{hajicek2001}. They describe the quantum evolution of a wave packet corresponding to a collapsing null shell of matter and also conclude that the shell undergoes a bounce rather than forming a singularity.

In 2014, Rovelli and Haggard presented an explicit metric describing this transition as observed from outside~\cite{haggard2014}. Notably, this metric provides an exact solution to the Einstein equations, except within a small, compact region of spacetime where the equations are violated.

In 2018, Ashtekar, Olmedo, and Singh analyzed quantum evolution within an eternal black hole \cite{ashtekar2018a}. They foliated the black hole’s interior with spacelike hypersurfaces and approximated quantum dynamics using an effective evolution similar to loop quantum cosmology (LQC). As a result, the classical singularity is replaced by a smooth transition surface, bridging a past trapped region with a future anti-trapped region.

Together, these results support the plausibility of a black-to-white hole transition. However, the interplay between this transition and Hawking radiation initially remained unclear. Haggard and Rovelli~\cite{haggard2014} estimated the \textit{bounce time}—the black hole lifetime before the transition—via dimensional analysis, concluding it to be of order \( M^2 \), which is shorter than the Hawking evaporation timescale of order \( M^3 \). Based on this, they suggested that evaporation could be neglected in the transition process.

In contrast, Frolov and Vilkovisky~\cite{frolov1979} estimated a bounce time of order \( M e^M \), which exceeds the evaporation time. They concluded that Hawking evaporation dominates the dynamics initially, and that the bounce occurs only at the final stage of black hole evaporation.

Loop quantum gravity (LQG), in its spinfoam formulation, offers a formalism for estimating the bounce time, though practical computation remains challenging without drastic approximations. Using this approach, Christodoulou and D’Ambrosio estimated the bounce time as \( M e^{\xi M^2} \)~\cite{christodoulou2018a,christodoulou2018b,dambrosio2020}, with \( \xi \) a positive constant, reinforcing the conclusion that evaporation cannot be neglected.

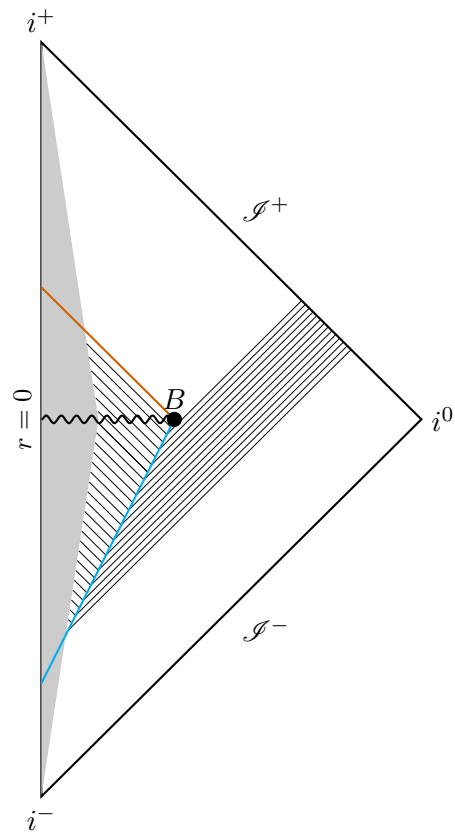
\begin{figure}[h!]
	\centering
	\begin{tikzpicture}[scale=5]
		\coordinate (O) at ( 0, 0);
		\coordinate (S) at ( 0,-1); 
		\coordinate (N) at ( 0, 1); 
		\coordinate (E) at ( 1, 0); 
		\coordinate (B) at ( 0.35, 0);
		\draw[thick] (N) -- (E) -- (S) -- cycle;
		
		\newcommand{\scri}{\mathscr{I}} 
		\node[right] at (E) {$i^0$};
		\node[above] at (N) {$i^+$};
		\node[below] at (S) {$i^-$};
		\node[above] at (B) {$B$};
		\node[above, rotate=90] at (O) {\small{$r=0$}};
		\node[above right] at (0.5,0.5) {$\scri^+$};
		\node[below right] at (0.5,-0.5) {$\scri^-$};
		
		\fill[gray!40] (0.15,0) -- (S) -- (N) -- cycle;
		
		\draw[thick,decorate,decoration={snake, amplitude=0.5mm, segment length=2mm}] 
		(0,0) to[out=0, in=180] (B);
		
		\foreach \x in {0.07,0.09,...,0.34} {
			\draw[line width=0.2] (\x, 2*\x - 0.7) -- ++(0.85 - 1.5*\x,0.85 - 1.5*\x);
		}
		
		\foreach \x in {0.08,0.09,...,0.28} {
			\draw[line width=0.2] (\x, 2*\x - 0.7) -- ++(-0.61* \x + 0.04, -0.04 + 0.61 * \x);
		}
		
		\foreach \x in {0.28,0.29,...,0.35} {
			\draw[line width=0.2] (\x, 2*\x - 0.7) -- ++(-1.59* \x + 0.32 , - 0.32 +  1.59 * \x);
		}
		
		\draw[thick,cyan] (0,- 0.7) -- (B);
		
		\draw[thick,orange!80!black] (0,0.35) -- (B);
		
		\fill[black] (B) circle (0.02);
	\end{tikzpicture}
	\caption{Penrose diagram of an evaporating black-to-white hole. The black hole forms from the collapse of a star (shaded region). Hawking pairs are produced along the time-like apparent horizon (blue line). In the hatched regions, each line represents either an ingoing or outgoing quantum. These regions correspond to Schwarzschild metrics with decreasing Misner–Sharp masses, serving as a first-order approximation of the radiation's backreaction on the geometry, following the approach introduced by Hiscock~\cite{hiscock1981,hiscock1981a}. Radiation falling into the black hole interacts with the collapsing matter or crosses the would-be singularity and enters an anti-trapped region. Beyond the white hole horizon (orange line), the metric becomes Schwarzschild with Planck-scale mass. The collapsing matter also traverses the would-be singularity and re-emerges. Point \( B \) marks a compact, Planck-scale region where the classical metric breaks down.}
	\label{fig:B2W}
\end{figure}

In this refined scenario, Hawking evaporation dominates the initial phase, gradually reducing the black hole’s mass. Once it reaches the Planck scale, the black hole tunnels into a white hole. This version of the scenario was first examined by Bianchi et al.~\cite{bianchi2018}. In~\cite{martin-dussaud2019a}, we further developed this model and studied the effective spacetime geometry resulting from the collapse of a null shell, including the backreaction of Hawking radiation on the metric. Figure~\ref{fig:B2W} shows a Penrose diagram of the process.

\section{An information-theoretic argument places a lower bound of \( M^4 \) on the lifetime of remnants}

In the black-to-white hole scenario, the transition occurs when the black hole has typically reached Planckian mass. Immediately after this transition, the resulting white hole must have the same mass. However, due to the large number of emitted Hawking quanta, the white hole remains highly entangled with the exterior. In other words, it still `contains a lot of information', encoded in its quantum correlations with the radiation field. This situation is characteristic of a \textit{long-lived remnant}—a generic term for whatever might persist if Hawking evaporation were to halt at late times, when semi-classical approximations cease to be valid.

The possibility of such remnants has been considered in early works, primarily to argue against it. For instance, in 1975, Hawking claimed that `\textit{because black holes can form when there was no black hole present beforehand, CPT implies that they must also be able to evaporate completely; they cannot stabilize at the Planck mass, as has been suggested by some authors}'~\cite{hawking1976}. However, this argument is flawed: black hole evaporation is not the time-reversal of gravitational collapse. Rather than implying complete evaporation, time-reversal symmetry suggests the possible existence of white holes.

In 1982, Hawking offered another argument against the existence of remnants, stating that `\textit{otherwise one would expect the mass density of the universe to be dominated by remnant black holes which would give rise respectively to a very large positive deceleration parameter}'~\cite{hawking1982}. However, he provides no justification for why remnants should be produced in such large numbers as to significantly affect the cosmic expansion.

In 1987, Aharonov, Casher, and Nussinov explored the concept of Planck-scale remnants, which they referred to as `planckons'~\cite{aharonov1987}, and argued that such objects, if they exist, would be stable. In the same year, Carlitz and Willey provided an estimate for the lifetime of these remnants~\cite{carlitz1987}. A more concise and influential argument, that I reproduce below, was later presented by Preskill~\cite{preskill1993}.

Over its lifetime, a black hole emits a total number \( N \) of Hawking quanta. Since the total radiated energy equals the initial mass \( M \), the average energy per quantum is \( E = M/N \). Moreover, Wien's displacement law states that the average energy is proportional to the temperature \( T \), which, according to Hawking's formula, scales as \( T \sim 1/M \). It follows that
\begin{equation}
	N \sim M^2.
\end{equation}

To preserve unitarity, a Planck-scale remnant must eventually emit at least \( N \) quanta, each carrying the missing information. However, the total available energy is now only of order unity (in Planck units), so the average energy per quantum must be \( \sim 1/M^2 \), corresponding to a wavelength of order \( M^2 \). 

Due to the monogamy of entanglement, these late-time quanta cannot be strongly correlated with each other, since they are already entangled with the earlier Hawking radiation. So each must be emitted independently rather than altogether. The emission for each quantum requires about the time for the wave to move out, i.e. \( M^2 \) in Planck units. Consequently, the total emission time, and hence the lifetime of the remnant, satisfies the lower bound
\begin{equation}
	T_{\text{remnant}} \geq M^4.
\end{equation}

This result has been incorporated into the black-to-white hole scenario, where the remnant is a white hole formed from the quantum decay of a black hole~\cite{bianchi2018}. In that context, the lower bound on the remnant lifetime has often been assumed to be saturated, without detailed justification. According to the resulting picture, the black hole undergoes Hawking evaporation for a time of order \( M^3 \), after which it tunnels into a Planck-scale white hole. The white hole then persists for a time of order \( M^4 \), gradually evaporating through the thermal emission of low-energy quanta. Although both phases of radiation are separately thermal, correlations between the early and late emissions encode the necessary information to preserve unitarity.

While this estimate may serve as a provisional benchmark, it is notably weak. In particular, it does not arise from a detailed analysis of the dynamical evolution of the system, but solely from the requirement that information be restored. Thus, it is not specific to white holes and could apply to any long-lived remnant. Preskill’s heuristic argument merely states that, if the quantum state of a remnant is to be purified, the process must take at least a time of order \( M^4 \). In the final section, I present a more robust argument based on the spacetime dynamics of the black-to-white hole transition, leading to a refined estimate of order \( M^5 \).

\section{The dynamical evolution of white hole interiors suggests a lifetime of \( M^5 \)}

White hole remnants exhibit a peculiar geometry: while their radius is Planck-scale, their interior volume is large. This structure is inherited from the late-stage black hole geometry just before tunneling. What changes across the transition is the dynamical behavior of the geometry: in the black hole phase, the interior volume grows over time (Schwarzschild like metric) while the radius decreases due to Hawking evaporation; in the white hole phase, the volume shrinks (Schwarzschild like metric) while the radius remains fixed at the Planck scale. The volume evolution is time-reversed, but the radius evolution is not, since it is governed by Hawking radiation—an inherently irreversible process. Notice that this asymmetry implies that two black holes of the same mass can be distinguished by their interior volume, which is larger for an older black hole.

In 2016, Christodoulou and De Lorenzo showed that the interior volume of a black hole with initial mass \( M \) evolves as
\begin{equation}
	\label{eq:interior-volume}
	V(T) \sim T M^2,
\end{equation}
where \( T \) is the retarded time since the black hole's formation~\cite{christodoulou2016a}. Although they also computed corrections due to Hawking evaporation, these do not affect the leading-order estimate. By the end of the evaporation process, \( T \sim M^3 \), yielding an interior volume \( V \sim M^5 \). The resulting white hole remnant thus inherits a volume of order \( M^5 \) at the moment of tunneling.

The metric of a white hole is the time-reverse of that of a black hole. Since equation~\eqref{eq:interior-volume} follows solely from the spacetime metric, it can be applied analogously to the white hole phase. In particular, the time required for a Planckian white hole with interior volume \( V \) to disappear is the same as the time it would take a Planckian black hole to develop an interior volume \( V \). Applying this reasoning to a remnant with volume \( V \sim M^5 \), we find that its lifetime is
\begin{equation}
	T_{\text{WH}} \sim \frac{V}{m_P^2} \sim M^5,
\end{equation}
where \( m_P \) is the Planck mass.

First, note that \( T_{\text{WH}} \geq M^4 \), making this estimate compatible with Preskill’s argument: it allows sufficient time for purification of the quantum state. Second, unlike purely information-theoretic bounds, this argument is grounded in the dynamical evolution of space, making it physically more compelling. Heuristically, the white hole takes a long time to evaporate because it must release a large interior volume through a tiny Planck-scale horizon. Finally, this longer lifetime relaxes phenomenological constraints on the hypothesis that white hole remnants could constitute dark matter~\cite{rovelli2024}.

\section*{Acknowledgments}

I am grateful to Carlo Rovelli for valuable discussions on this topic, and to Alexandra Elbakyan for her help in accessing scientific literature. This work is licensed under a Creative Commons Attribution 4.0 International License (CC BY 4.0).

\bibliographystyle{quantum}
\bibliography{m5}

\end{document}